\documentclass[%
twocolumn,
 amsmath,amssymb,
 aps, 
pra,
]{revtex4-2}

\usepackage{graphicx}
\usepackage{dcolumn}
\usepackage{bm}
\usepackage{ulem}
\usepackage{xcolor}

\usepackage[colorlinks=true, allcolors=blue]{hyperref}

\begin{document}

\title{\textbf{Parametric dependence of unidirectional guided resonances in periodic structures} 
}%

\author{Lijun Yuan}
\email{ljyuan@ctbu.edu.cn}
\affiliation{
 School of Mathematics and Statistics, Chongqing Technology and Business University, Chongqing 400067, China
}%

\author{Ya Yan Lu}
\affiliation{%
 Department of Mathematics, City University of Hong Kong, Kowloon, Hong Kong, China
}%

\begin{abstract}
Unidirectional guided resonances (UGRs) in periodic structures are special resonant modes that exhibit strict one-sided radiation, even though radiation in both sides is allowed, offering significant advantages for various applications.  Under a structural perturbation, a UGR typically turns to a regular resonant mode that radiates to both sides.  Existing numerical results indicate that to find UGRs in any periodic structure, it is necessary to tune at least one parameter. In this work, we develop a rigorous theory on the parametric dependence of UGRs. {We show that in the presence of a single radiation channel, a UGR can exist continuously with respect to a structural parameter, provided that another parameter (associated with a generic perturbation) is properly tuned.} Moreover, from a periodic structure with a {generic} bound state in the continuum (BIC), it is always possible to obtain a continuous family of UGRs by tuning one parameter. This implies that UGRs with arbitrarily large quality factor can be easily obtained.  Our work provides a theoretical basis for designing useful photonic devices based on UGRs.
\end{abstract}

\maketitle

\section{Introduction}

Unidirectional guided resonances (UGRs) in periodic structures, such as photonic crystal (PhC) slabs, are special resonant modes that radiate power only to one side of the structure, even though radiation in both sides is allowed \cite{zhou16,yin20,zeng21}. Compared to other methods for achieving unidirectional radiation, such as using mirrors \cite{zhu17} or substrate reflectors \cite{sub15}, UGRs in PhC slabs offer a more compact and integrable solution for photonic platforms. This makes UGRs attractive for various applications, including all-pass phase shifters \cite{zhang22}, low-loss optical interconnects \cite{wang24}, and intensity-flattened phase modulation devices \cite{zhang21,xu23}.

In a periodic structure  with no up-down mirror symmetry, resonant modes naturally exhibit asymmetric radiation.  This can be quantified by the asymmetry ratio defined as the ratio of power radiated to the top and the bottom. For practical applications, it is of significant interest to design structures that exhibit large asymmetry ratios \cite{wang13,ota15}. A UGR corresponds to a resonant mode with a zero or infinite asymmetry ratio.  A fundamental theoretical question is: Can this ratio truly reach zero or infinity?  In \cite{wang13},  theoretical bounds for the asymmetry ratio were derived using a temporal coupled-mode theory, and it was shown that the lower and upper bounds can reach zero and infinity, respectively. Later, Zhou {\it et al.} showed that breaking the up-down mirror symmetry of a PhC slab with a bound state in the continuum (BIC) enables the realization of zero or infinite asymmetry ratios \cite{zhou16}. Based on this method, UGRs were experimentally realized in \cite{yin20}. Note that near a BIC, the quality factor ($Q$ factor) of UGRs can be arbitrarily large \cite{xu23b}. In addition to the BIC-based method, UGRs can also be realized through other mechanisms \cite{peng18,muk21,song22,han23,yin23,lee24,zheng25}. All existing approaches for achieving UGRs rely on tuning structural parameters. Therefore, it is important to find out how UGRs depend on parameters.   Existing numerical results indicate that typical UGRs form a continuous curve in the plane of two structural parameters \cite{yin20,xu23b}. {In addition, it is easy to show that for any UGR with a single radiation channel,   the inverse of the corresponding $2\times 2$ scattering matrix must have a zero column. This suggests that to find a URG, it is necessary to tune one structural parameter. However, currently, a rigorous, general and quantitative theory is still lacking. }

In this paper, we present a {formal} study on how UGRs in periodic structures depend on generic structural parameters. More precisely, we show that in the simplest case with a single radiation channel, generic UGRs exist as a curve in the plane of two parameters,  and more generally, they form a codimension-1 object in parameter space. {In addition, from   a generic bound state in the continuum (BIC) in a periodic structure , it is always possible to obtain a continuous family of UGRs by tuning one parameter.} We establish this result using a perturbation method which constructs the UGRs explicitly as parameters are varied. {Our method provides a quantitative analysis with explicitly computable expansion coefficients, and  uncovers the precise conditions (on both the UGR and the perturbation) under which   the codimension-1 property holds.} 
The perturbation method has previously been used to study the parametric dependence of BICs \cite{yuan20}.  To extend the method to UGRs, two new challenges must be addressed. Firstly, the method requires a diffraction solution at the frequency of the UGR, but the resonant frequency is a pole of the scattering matrix, and it is not clear whether diffraction solutions at the complex resonant frequency can be defined. Secondly, since UGRs are resonant  modes, they are not square-integrable. These difficulties make the extension of perturbation theory from BICs to UGRs nontrivial. We show that our perturbation method requires the diffraction solution at the complex resonant frequency for a specific incident wave and it is well-defined. To address the integrability issue, we use the perfectly matched layer (PML) regularization technique \cite{lala13,lala22}. The rest of this paper is organized as follows. In Sec.~\ref{sec:uni_resonance}, we present the mathematical formulations for UGRs, construct the appropriate diffraction solution, and introduce the PML regularization technique. In Sec.~\ref{sec:perturbation}, we present the perturbation method to construct the UGRs. Numerical examples are presented in Sec.~\ref{sec:numerical}. The paper is concluded with some remarks in Sec.~\ref{sec:conclusions}.

\section{Unidirectional guided resonances, diffraction solutions and Regularization}
\label{sec:uni_resonance}
\subsection{Unidirectional guided resonances}
Consider a two-dimensional (2D)  structure   that is  invariant in $x$, periodic in $y$ with period $L$, and sandwiched between two  homogeneous media in the $z$ direction. The dielectric function $\epsilon(y,z)$ satisfies $\epsilon(y,z) = \epsilon(y+L,z)$ for all $(y,z)$, and $\epsilon(y,z) = \epsilon_0$ for $|z| > d$, where $2d$ is the thickness of the periodic layer.  For the $E$-polarization, the $x$-component of the electric field, denoted as $u$, satisfies the 2D Helmholtz equation:
\begin{equation}
    \label{helm} \frac{\partial^2 u}{\partial y^2} + \frac{\partial^2 u}{\partial z^2} + k^2 \epsilon(y,z) u = 0,
\end{equation}
where $k = \omega/c$ is the free-space wave number, $\omega$ is the angular frequency, and $c$ is the speed of light in vacuum. A UGR is a special resonant mode with a complex frequency that  radiates  power only to one side of the structure. Without loss of generality, we consider UGRs that radiate only downward, that is, they tend to zero as $z \to +\infty$ and satisfy an outgoing radiation condition as $z \to -\infty$.

For simplicity,  we consider resonant modes with only one radiation channel on each side, that is, the Bloch wave number $\beta$ and frequency $\omega$ satisfy 
\begin{equation}
\label{one_channel}
\beta^2 < \mbox{Re} \left(k^2 \epsilon_0 \right) < \left(  2\pi /L - |\beta|\right)^2.
\end{equation}
Since the structure is periodic and homogeneous for $|z| > d$, any resonant mode $u(y,z)$ can be expanded in Fourier series as
\begin{equation}
\label{fourier}
u(y,z) = \sum\limits_{j=-\infty}^{\infty} c_j^{\pm} e^{ {\sf i} [\beta^{(j)} y \pm \gamma^{(j)} (z \mp d) ]}, \quad \pm z  >d,
\end{equation}
where $c_{j}^{\pm}$ are the Fourier coefficients of $u(y,\pm d)$, $\beta^{(j)} = \beta + 2\pi j/L$, $\gamma^{(j)} = \sqrt{k^2 \epsilon_0 - \left[\beta^{(j)}\right]^2}$ ({the square root is defined with a branch cut along the negative imaginary axis}), and $c_0^{\pm}$ are the upward and downward radiation coefficients of the resonant mode. With condition (\ref{one_channel}),  a resonant mode becomes a UGR that radiates only downward if and only if $c_0^+=0$ and $ c_0^- \neq 0$. The asymmetry ratio is defined as $\tau = |c^+_0/c^-_0|^2$, and a UGR corresponds to $\tau = 0$. 

Numerical results indicate that under a general structural perturbation, a UGR will become a regular resonant mode which radiates to both sides of the structure \cite{yin20}. This does not mean that the UGR cannot be preserved if the perturbation contains tunable parameters. We want to know how many parameters must be tuned in order to maintain the continual existence of a UGR under a general structural perturbation. In the following, we develop a theory on the parametric dependence of UGRs, and show that one tunable parameter {associated with a generic perturbation} is sufficient {to ensure the continual existence of}  UGRs with a single radiation channel. 

To develop the theory on UGRs, we denote the dielectric function of a periodic structure with a UGR by $\epsilon_*(y,z)$, denote the UGR by $u_*(y,z)$, its frequency by $\omega_*$, and its Bloch wave number by $\beta_*$, where $\omega_*$ and $\beta_*$ satisfy condition (\ref{one_channel}). In addition, we denote the upward and downward radiation coefficients of the UGR  by $c^+_*$ and $c^-_*$, respectively, where $c^+_* = 0$ and $c^-_* \neq 0$. {By reciprocity, there exits a resonant mode $v_*(y,z)$ for the same resonant frequency $\omega_*$ and opposite Bloch wave number $-\beta_*$. We call $v_*$  the reciprocal mode of $u_*$.} If the structure has no reflection symmetry in $y$ and $\beta_* \neq 0$, then  $v_*$ is typically not a UGR. Otherwise,  {as shown in Appendix A}, it is also a UGR that radiates to the same side.   Let $d^{\pm}_*$ be the upward and downward radiation coefficients of the {reciprocal} mode, we scale $u_*$ and  $v_*$ such that $|c^-_*|^2 = |d_*^+|^2 + |d_*^-|^2$.

\subsection{Diffraction solutions}
Our theory is developed using a perturbation method that constructs the UGR in perturbed structure explicitly. 
In this method,  we need a diffraction solution at the complex resonant frequency $\omega_*$ of the original UGR in the unperturbed structure. 
The solution of a diffraction problem can be described by a  scattering matrix, and the resonant frequencies are the poles of the determinant of the scattering matrix \cite{popov86, wu22}. For a regular resonant mode, the resonant frequency is also a pole of each element of the scattering matrix, which implies that the associated diffraction problem has no bounded solution at the complex resonant frequency. However, as shown in Appendix A, if  there is a UGR at  $(\beta_*, \omega_*)$, the  diffraction problem at $\omega_*$ is well-defined for incident waves with Bloch wave number $-\beta_*$. {    More specifically, using Fredholm alternative, we show that
if the UGR is nondegenerate, then for an incident wave coming from the top with $(-\beta_*, \omega_*)$, i.e.,  $u^{(\sf in)} = e^{- {\sf i} [\beta_* y + \gamma_* (z - d)]}$ for $ z >d$, where $\gamma_* =  \sqrt{k_*^2 \epsilon_0 - \beta_*^2}$,   the diffraction problem is well-defined.} The corresponding solution $w_*(y,z)$ has the following asymptotic relation
\begin{equation}
\label{asymp_diff} w_*  \sim  e^{- {\sf i} [\beta_* y + \gamma_* (z - d)]} + r_* e^{- {\sf i} [\beta_* y - \gamma_* (z - d)]}, \quad z \to +\infty,
\end{equation}
where {$r_* $ is the reflection coefficient.}

 \subsection{Regularization}
Since a UGR diverges as $z \to -\infty$, it is not square-integrable over one period of the structure $\Omega = \left\{ (y,z) : \ |y|<L/2, |z| < +\infty \right\}$.   To overcome this difficulty, we use the perfectly matched layer (PML) regularization technique \cite{lala13,lala22}. Since we consider UGRs that radiate downward, we place a PML in the negative $z$ direction. The PML is equivalent to a complex coordinate stretching  $\hat{z} = \int_{0}^z s(\zeta) d \zeta$ \cite{chew94}, where $s(z) = 1 + {\sf i} \mu $ for $z< - h$ ($\mu > 0$, $h > d$) and $s(z) = 1$ otherwise. Note that outside the PML, we have $\hat{z} = z$ for $z > -h$. With the PML, the $x$-component of the electric field, $u(y,\hat{z})$,  satisfies
\begin{equation}
    \label{helm_pml} 
    \frac{\partial^2 u(y, \hat{z}) }{\partial y^2}  + \frac{\partial^2 u(y, \hat{z})}{\partial \hat{z}^2}  + k^2 \epsilon(y, \hat{z}) u(y, \hat{z}) = 0,
\end{equation}
where $\epsilon(y, \hat{z}) = \epsilon(y,z)$ for all $(y,z) \in \Omega$.
The PML affects only the domain for $z < -h$. Therefore, $u(y, \hat{z})$ is identical to the original $u(y,z)$ for $z > -h$, and has the same asymptotic relation as $z \to +\infty$ and the same Fourier expansion (\ref{fourier})  for $z > -h$.  With the PML,  a UGR that radiates downward decays to zero as $z \to -\infty$ and its zeroth-order Fourier coefficient $c_0^-  \neq 0$. 

{Let $u_*(y, \hat{z})$,  $v_*(y,\hat{z}) $,  $w_*(y, \hat{z}) $ denote the UGR, the {reciprocal}  mode and the associated diffraction solution when the PML is introduced, respectively, then $v_*(y,\hat{z}) $ is the reciprocal mode of $u_*(y, \hat{z})$.}   To ensure that $u_*(y, \hat{z})$ and  $v_*(y,\hat{z}) $ decay to zero as $z \to -\infty$, the constant $\mu$ should satisfy $\mu \mbox{Re}(\gamma_*) + \mbox{Im}(\gamma_*) > 0$. Equation (\ref{asymp_diff}) is still valid for $w_*(y, \hat{z}) $.  With the PML, we can normalize the UGR and the  {reciprocal   mode}. We assume 
\begin{equation}
\label{generic1}
\int_{\hat{\Omega}} v_*(y,\hat{z}) \epsilon_*(y,\hat{z})   u_*(y,\hat{z}) dx d\hat{z} \neq 0,
\end{equation}
where $\hat{\Omega} = (-L/2, L/2) \times \Gamma$, and 
$$\Gamma = \left\{  \hat{z} \in \mathbb{C} : \ \hat{z} = \int_0^z s(\zeta) d\zeta, \, z \in \mathbb{R} \right\}$$
is an infinite curve in the complex plane. Under assumption (\ref{generic1}), we can choose the amplitudes of $u_*$ and $v_*$, and the phase of $v_*$ such that 
\begin{equation}
    \label{normlization}
\int_{\hat{\Omega}} v_*(y,\hat{z}) \epsilon_*(y,\hat{z})   u_*(y,\hat{z}) dx d\hat{z} =1.
\end{equation}
The existence of the {reciprocal mode $v_*(y,\hat{z}) $} implies that the diffraction solution  is not unique. That is, if $w_*(y,\hat{z})$  is a solution, then $w_*(y,\hat{z}) + \xi v_*(y,\hat{z})$ for any constant $\xi$ is also a solution for the same incident wave. Therefore, we can choose an appropriate $\xi$ such that 
\begin{equation}
\label{orth_w}
\int_{\hat{\Omega}} w_*(y,\hat{z}) \epsilon_*(y,\hat{z}) u_*(y, \hat{z}) dyd\hat{z}  = 0. 
\end{equation}
In addition, to simplify the perturbation analysis in the following section, we choose the phase of $u_*$ such that the integral
\begin{equation}
    \label{real_u}
     \int_{\hat{\Omega}} w_*(y, \hat{z})  \partial_y u_*(y, \hat{z}) dyd\hat{z}
\end{equation}
is purely  imaginary.

For simplicity, we write the UGR, the diffraction solution, and the {reciprocal} mode as $u_* = \phi_* e^{ {\sf i} \beta_* y}$,  $w_* = \varphi_* e^{- {\sf i} \beta_* y}$ and  $v_* = \psi_* e^{- {\sf i} \beta_* y}$, respectively, where $\phi_*, \varphi_*$ and $\psi_*$ are periodic functions of $y$ with period $L$.  Clearly, $\phi_*$ satisfies the following  equation
\begin{equation}
    \label{helm_phi}  \frac{\partial^2 \phi_* }{\partial y^2}  + \frac{\partial^2 \phi_*}{\partial \hat{z}^2} + 2 {\sf i} \beta_* \frac{\partial \phi_*}{\partial y}   + \left(k^2_* \epsilon_* - \beta^2_* \right) \phi_* = 0,
\end{equation}
and $\psi_*$ and $\varphi_*$ satisfy the same equation with $\beta_*$  replaced by $-\beta_*$.
In Sec.~\ref{sec:perturbation}, the theory is developed  using $\phi_*$ instead of $u_*$. Corresponding to Eqs.~(\ref{asymp_diff}), (\ref{normlization}), (\ref{orth_w}) and (\ref{real_u}), we have the following relations
\begin{eqnarray}
\label{asymp_diff2} 
 & & \varphi_*(y,\hat{z})  \sim  e^{- {\sf i}  \gamma_* (z - d)} + r_* e^{ {\sf i} \gamma_* (z - d)},  \quad z \to +\infty, \\
 \label{normalization2} 
 && \int_{\hat{\Omega}} \psi_*(y,\hat{z}) \epsilon_*(y,\hat{z}) \phi_*(y, \hat{z}) dyd{\hat{z}}  = 1, \\
\label{orth}
& & \int_{\hat{\Omega}} \varphi_*(y,\hat{z}) \epsilon_*(y,\hat{z}) \phi_*(y, \hat{z}) dyd{\hat{z}}  = 0,
\end{eqnarray}
and the integral
\begin{equation}
    \label{real}
     \int_{\hat{\Omega}} \varphi_*(y, \hat{z}) \left[\beta_* \phi_*(y, \hat{z}) - {\sf i} \partial_y \phi_*(y, \hat{z}) \right]dyd{\hat{z}}
\end{equation}
is real.

\section{Parametric dependence theory}
\label{sec:perturbation}
In this section, we present our theory on parametric dependence of  UGRs  using a perturbation method. Let $\epsilon_*$ be the dielectric function of an unperturbed periodic structure with a UGR. If the perturbed structure has a dielectric function given by
\begin{equation}
\label{perturbation1}
\epsilon(y,z) = \epsilon_*(y,z) + \delta F(y,z),
\end{equation}
where $F$ is the perturbation profile and $\delta$ is the amplitude of the perturbation, then in general, the UGR becomes a regular resonant mode which radiates to both sides of the structure. We want to show that if an additional parameter is introduced in the perturbation, then the UGR can exist continuously with respect to $\delta$. More precisely, we consider the dielectric function 
\begin{equation}
\label{perturbation}
\epsilon(y,z) = \epsilon_*(y,z) + \delta F(y,z) + \eta G(y,z),
\end{equation}
where $G$ is an independent perturbation profile, and $\eta$ is a small real tunable parameter. We assume that both $F$ and $G$ are {\sout{real}} $O(1)$ functions, periodic in $y$ with period $L$, and vanish when $|z| > d$. Thus, the perturbation preserves the periodicity and does not affect the surrounding homogeneous media. In addition,  $G$ must satisfy a generic condition to be specified below.

Our main result is as follows: If in the unperturbed structure $\epsilon_*$, there is a nondegenerate UGR $u_*(y,z) = \phi_*(y,z) e^{ {\sf i} \beta_* y}$  satisfying conditions (\ref{generic1}) and (\ref{generic_cond_phi}),  then for any real $\delta$ near $0$, there exists a real $\eta$ near $0$ such that the perturbed structure with dielectric function given in Eq.~(\ref{perturbation}) has a UGR $u(y,z) = \phi(y,z) e^{ {\sf i} \beta y}$ with $\beta$ close to $\beta_*$ and $\omega$ close to $\omega_*$. 
The above proposition implies that a generic UGR [i.e., one satisfies conditions (\ref{generic1}) and (\ref{generic_cond_phi})] exists as a curve in the parameter space $\delta$-$\eta$.

To prove the proposition, we use a perturbation method to construct the UGR in the perturbed structure. We show that for a properly chosen $\eta$ (depending on $\delta$), there is a UGR $u(y,z) = \phi(y,z) e^{ {\sf i} \beta y}$  in the perturbed structure for any small $\delta$.  To construct the UGR,  we expand $\phi(y,\hat{z}), \beta, k^2 = (\omega/c)^2$ and $\eta$ as power series of $\delta$:
\begin{eqnarray}
    \label{exp_phi}
    \phi(y,\hat{z}) &=& \phi_*(y, \hat{z}) + \delta \phi_1(y, \hat{z}) + \delta^2 \phi_2(y, \hat{z}) + \ldots, \\
    \label{exp_beta}
    \beta &=& \beta_* + \delta \beta_1 + \delta^2 \beta_2 + \ldots, \\
    \label{exp_k2} 
    k^2 &=& k_*^2 + \delta g_1 + \delta^2 g_2 + \ldots, \\
    \label{exp_eta}
    \eta &=&  \delta \eta_1 + \delta^2 \eta_2 + \ldots .
\end{eqnarray}
In the following, we show that for each $j \geq 1$,  $\beta_j$, $g_j$ and $\eta_j$ can be determined, $\beta_j$ and $\eta_j$ are real, and $\phi_j(y,\hat{z})$ can be solved and it decays to zero as $z \to +\infty$. Note that $g_j$ is typically not real, since a UGR has a complex frequency.

Substituting the expansions (\ref{exp_phi}) - (\ref{exp_eta}) into the governing equation of $\phi$, i.e., Eq.~(\ref{helm_phi}) with $\epsilon_*$ and $k_*$ replaced with $\epsilon$ and $k$, respectively, comparing the coefficients of $\delta^j$ for $j \geq 1$, we obtain the governing equation of $\phi_j$ as
\begin{equation}
    \label{eq_phij}
   \hat{ \mathcal{L}}_* \phi_j = \beta_j B_1 + g_j B_2 + \eta_j B_3 - C_j,
\end{equation}
where the operator $\hat{\mathcal{L}}_*$, and functions $B_1$, $B_2$, $B_3$ and $C_j$ are given in Appendix B. Note that $C_j$ only involves $\beta_m, g_m, \eta_m$ and $\phi_m(y, \hat{z})$ for $m \leq j-1$.  

To ensure Eq.~(\ref{eq_phij}) has a solution that decays to zero as $z \to +\infty$, $\beta_j, g_j$ and $\eta_j$ must satisfy  the following $2 \times 3$ linear system 
\begin{equation}
    \label{system_betaj}
    \begin{bmatrix}
        a_{11} & -1 & a_{13} \\
        a_{21} & 0 & a_{23} 
    \end{bmatrix} \begin{bmatrix}
        \beta_j \\ g_j \\ \eta_j
    \end{bmatrix}  = \begin{bmatrix}
        b_{1j} \\ b_{2j}
    \end{bmatrix},
\end{equation}
where $a_{11}, a_{13}, a_{21}$, $a_{23}$, $b_{1j}$ and $b_{2j}$ are given in Appendix B. The above system is obtained by  multiplying $\varphi_*$ and $\psi_*$ to both sides of Eq.~(\ref{eq_phij}) and integrating on $\hat{\Omega} $. Note that elements $-1$ and $0$ in the coefficient matrix above are obtained from the normalization condition (\ref{normalization2}) and the orthogonal condition (\ref{orth}), respectively, and $a_{21}$ is real due to assumption (\ref{real}). 

To ensure linear system (\ref{system_betaj}) has a unique solution  with real $\beta_j$ and real $\eta_j$, it is necessary to impose the following conditions:  
\begin{equation}
\label{generic_cond_phi}
a_{21} =  \int_{\hat{\Omega}}\varphi_* \left(\beta_* \phi_* - {\sf i} \partial_y \phi_* \right) dyd{\hat{z}} \neq 0,
\end{equation}
\begin{equation}
\label{generic_cond}
\mbox{Im}(a_{23} ) = -\mbox{Im}\left(k_*^2 \int_{\hat{\Omega}} \varphi_*  G \phi_* dyd{\hat{z}} \right) \neq 0.
\end{equation}
In that case, real $\beta_j$ and $\eta_j$ can be solved from 
\begin{equation}
    \label{system_real}
    \begin{bmatrix}
        a_{21}  & \mbox{Re}(a_{23}) \\
         0 & \mbox{Im}(a_{23}) 
    \end{bmatrix} \begin{bmatrix}
        \beta_j \\ \eta_j
    \end{bmatrix}  = \begin{bmatrix}
        \mbox{Re}(b_{2j} ) \\  \mbox{Im}(b_{2j})
    \end{bmatrix} .
\end{equation}
After that, we can solve $g_j$ from the first equation of (\ref{system_betaj}).  Once $\beta_j, g_j$ and $\eta_j$ are solved, we can solve $\phi_j$ from Eq.~(\ref{eq_phij}) and it decays to zero as $z \to +\infty$.  Repeating this procedure for all $j \geq 1$, we obtain a UGR given by power series (\ref{exp_phi}).

The above procedure remains valid if the original UGR happens to be a BIC. This implies that a continuous family of UGRs can be obtained from a BIC by tuning a single structural parameter. Consequently, our theory also reveals the existence of UGRs near BICs. More precisely, if $u_*(y,z)$ is a generic BIC and the perturbation profile $G$ satisfies the condition (\ref{generic_cond}), then following the above procedure, we can construct a continuous family of UGRs depending on $\delta$. Note that if the unperturbed structure has the up-down mirror symmetry, the perturbation profile $F$ must breaks this symmetry. In that case, the coefficient $g_1$ in series (\ref{exp_k2}) is real, and the $Q$ factor of the UGRs is proportional to $1 / \delta^2$. Therefore, arbitrarily large $Q$ factor can be obtained if $\delta$ is sufficiently small.

\section{Numerical examples}
\label{sec:numerical}
To validate the theory presented above, we consider a periodic array of dielectric rectangular cylinders surrounded by vacuum. The widths of the rectangular cylinders in the $y$ and $z$ directions are $W_y = 0.55 L$ and $W_z = 0.6 L$, respectively, where $L$ is the period in the $y$ direction. The dielectric constant of the cylinders is $12$. The structure supports a symmetry-protected BIC with free-space wave number $k_* = 0.5143 (2\pi/L)$  and Bloch wave number $\beta_* = 0$, and a propagating BIC with $k_* = 0.5746 (2\pi/L)$ and  $\beta_* = 0.2255 (2\pi/L)$. 

\begin{figure}[thp]
    \centering
    \includegraphics[scale=0.7]{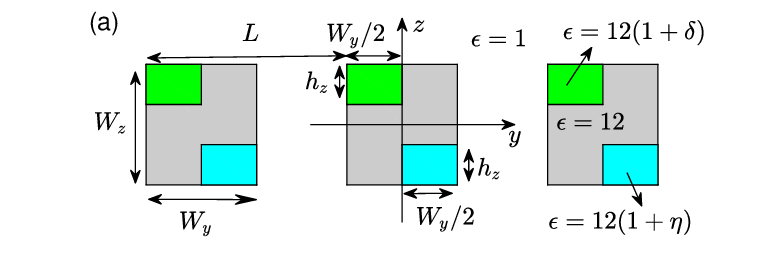}
     \includegraphics[scale=0.7]{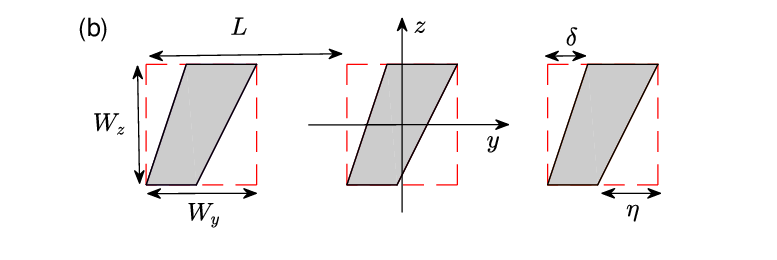}
    \caption{Structural perturbations to a periodic array of rectangular cylinders with period $L$ in the $y$ direction. (a): dielectric function perturbation. The dielectric constants of the upper-left and lower-right rectangles are changed to $12(1+\delta)$ and $12(1+\eta)$, respectively. (b): geometry perturbation. $\delta$ and $\eta$ are amplitudes of the perturbations. }
    \label{fig:structure}
\end{figure}

We consider two types of structural perturbations. The first perturbed structure is shown in Fig.~{\ref{fig:structure}}(a). The dielectric function $\epsilon(y,z)$  is given by Eq.~(\ref{perturbation})  with $F$ and $G$ defined as follows
\begin{eqnarray*}
F(y,z) &=& \left\{ \begin{matrix}
    12, &  (y,z) \in \left[-\frac{W_y}{2}, 0 \right] \times \left[ \frac{W_z}{2} - h_z, \frac{W_z}{2} \right], \\
    0,  &  \mbox{otherwise}, 
\end{matrix} \right. \\
G(y,z) &=& \left\{ \begin{matrix}
    12, &  (y,z) \in  \left[0, \frac{W_y}{2} \right] \times \left[ -\frac{W_z}{2}, -\frac{W_z}{2} + h_z \right], \\
    0,  &  \mbox{otherwise}
\end{matrix} \right.
\end{eqnarray*}
for $|y| < \frac{L}{2}$, where $h_z = 0.15 L$.  The second perturbed structure is shown in Fig.~{\ref{fig:structure}}(b). If $\delta \neq 0$, the perturbed structures are asymmetric in  the $z$ direction.  Our theory predicts that generic UGRs exist continuously with respect to $\delta$ for a properly chosen $\eta$ (depending on $\delta$). This implies that a family of  UGRs (depending on $\delta)$ can be obtained from a generic BIC.

\begin{figure}[thp]
    \centering
    \includegraphics[scale=0.65]{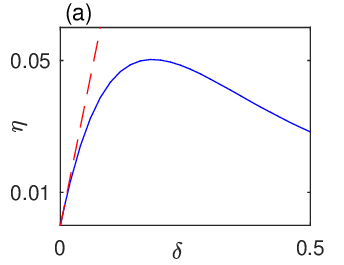} \qquad
     \includegraphics[scale=0.65]{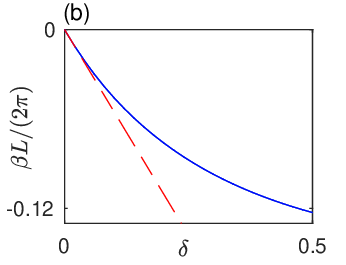} \\
     \begin{minipage}{0.5\linewidth}
     \centering
      \includegraphics[scale=0.65]{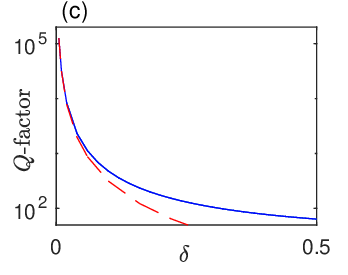}
      \includegraphics[scale=0.65]{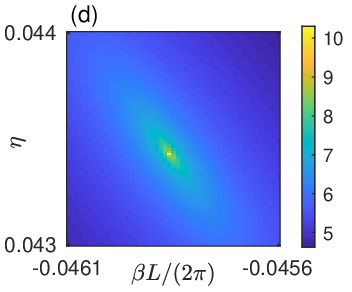}
    \end{minipage}
    \begin{minipage}{0.45\linewidth}
       \includegraphics[scale=0.65]{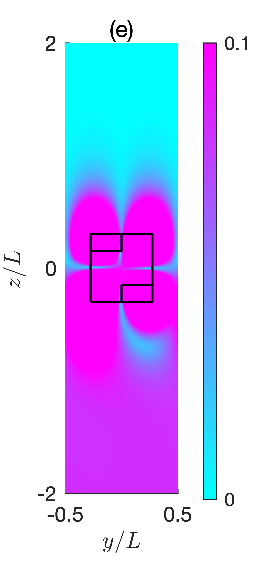}
       \end{minipage}
    \caption{Example 1: UGRs for the perturbation given in Fig.~\ref{fig:structure}(a) near the symmetry-protected BIC. (a): $\eta$ as a function of $\delta$; (b): $\beta$ as a function of $\delta$. Solid blue curves are the results of the FEM, and dashed red lines are the results of first-order approximation. (c): $Q$ factors of the UGRs for various $\delta$. The dashed red curve is the reference line for $ Q = 3.2/ \delta^2$. (d):  $\mbox{log}(1/\tau)$ for various $\eta$ and $\beta$ with $\delta = 0.1$. (e): the wave field amplitude $|u|$ of the UGR at $\delta = 0.1$. The color scale is limited to $[0, 0.1]$.}
    \label{fig:example1}
\end{figure}

In the first example, we consider the dependence of UGRs on the perturbation given in Fig.~\ref{fig:structure}(a) near the symmetry-protected BIC. Using the finite-element method (FEM), we calculat  UGRs that radiate only downward for various $\delta$. In Fig.~\ref{fig:example1}(a) and (b), the solid blue curves show the dependence of parameter $\eta$ and Bloch wave number $\beta$ (of the UGRs) on $\delta$, respectively. The first-order perturbation terms in Eqs.~(\ref{exp_beta}) and (\ref{exp_eta}) are $\eta_1 = 0.7486$ and $\beta_1 = -0.5522 (2 \pi /L)$, respectively. They are obtained by solving the linear system (\ref{system_real}) for $j=1$.  The first-order approximations, i.e. $\eta = \eta_1 \delta$ and $\beta = \beta_1 \delta$, are shown as red dashed lines in Fig.~\ref{fig:example1}(a) and (b). It can be observed that they are accurate for small $\delta$. In Fig.~\ref{fig:example1}(c), we show the $Q$ factor of the UGRs for various $\delta$. An extremely large  $Q$ factor can be obtained if $\delta$ is sufficiently close to $0$. For $\delta = 0.1$ and  $\eta = 0.0434$, there is a UGR with $\beta = -0.0459 (2\pi/L)$ and $\omega L / (2 \pi c) = 0.5071-5.29\times 10^{-4} {\sf i}$. For $\eta$ and $\beta$ near these values, we have regular resonant modes. In Fig.~\ref{fig:example1}(d), we show the logarithm of  the inverse asymmetry ratio of the resonant modes, i.e. $\mbox{log}(1/\tau) = \mbox{log}(|c_0^-/c_0^+|^2)$, for various $\eta$ and $\beta$.  We can see that the maximum numerical value of $1/\tau$ exceeds $10^{10}$.  This confirms the existence of a UGR that radiates only downward. The wave field amplitude (i.e., $|u|$) of the UGR for $\delta = 0.1$ is shown in Fig.~\ref{fig:example1}(e), where the color scale is limited to $[0, 0.1]$. It is evident that there is no radiation in the upward direction.

\begin{figure}[thp]
    \centering
    \centering
    
    \includegraphics[scale=0.65]{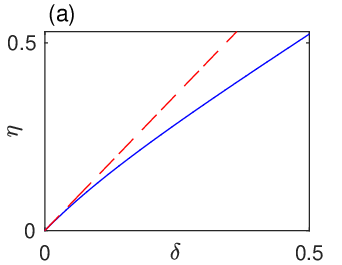} \qquad
     \includegraphics[scale=0.65]{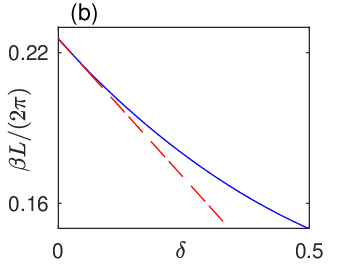} \\
     \begin{minipage}{0.5\linewidth}
      \includegraphics[scale=0.65]{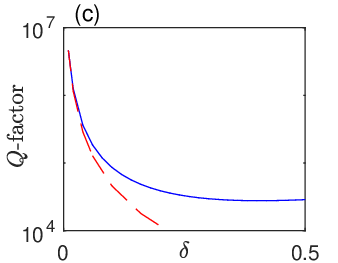}
      \includegraphics[scale=0.65]{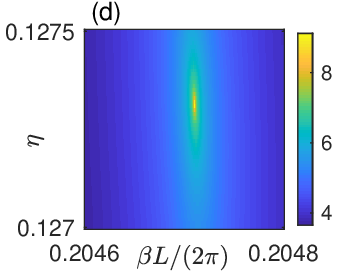}
    \end{minipage}
    \begin{minipage}{0.45\linewidth}
       \includegraphics[scale=0.65]{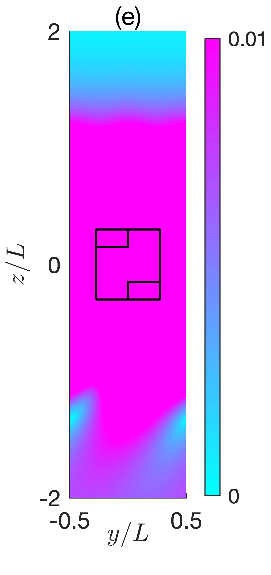}
       \end{minipage}
    \caption{Example 2: UGRs for the perturbation given in Fig.~\ref{fig:structure}(a) near the propagating BIC. (a): $\eta$ as a function of $\delta$; (b): $\beta$ as a function of $\delta$. Solid blue curves represent  the results of the FEM, and dashed red lines are the results of first-order approximation. (c): $Q$ factors of the UGRs for various $\delta$. The dashed red curve is the reference line for $Q = 459.6/ \delta^2$. (d): $\mbox{log}(1/\tau)$ for various $\eta$ and $\beta$ with $\delta = 0.1$. (e): the wave field amplitude of the UGR at $\delta = 0.1$. The color scale is limited to $[0, 0.01]$.}
    \label{fig:example2}
\end{figure}

In the second example, we study UGRs for the perturbation given in Fig.~\ref{fig:structure}(a) near the propagating BIC. A family of UGRs  for $\delta$  varying from $0$ to $0.5$ is found numerically. In Fig.~\ref{fig:example2}(a), we show $\eta$ depending on $\delta$ and the first-order approximation, i.e., $\eta = \eta_1 \delta$ where $\eta_1 = 1.4613$, as the blue solid and red dashed lines, respectively. It is clear that the first-order approximation is accurate for small $\delta$. Similarly, $\beta$ depending on $\delta$ and its first-order approximation $\beta = \beta_* + \beta_1 \delta$ where $\beta_1 = -0.2216(2 \pi /L)$, are shown as the blue solid and red dashed lines in Fig.~\ref{fig:example2}(b), respectively. The $Q$ factor of the UGR can be very large if $\delta$ is close to $0$ as shown in Fig.~\ref{fig:example2}(c). For $\delta = 0.1$ and $\eta = 0.1273$, there is a UGR with $\beta = 0.2047 (2\pi/L)$ and $\omega L / (2 \pi c) =  0.5655-3.28\times 10^{-6} {\sf i}$. Its wave field amplitude is shown in Fig.~\ref{fig:example2}(e). The color scale is limited to $[0, 0.01]$, and there is no radiation in the upward direction. For different $\eta$ and $\beta$ near those of the UGR, we show $\mbox{log}(1 / \tau)$  in Fig.~\ref{fig:example2}(d). The maximum numerically computed $1/\tau$ exceeds $10^{9}$, confirming the existence of the UGR.

\begin{figure}[thp]
    \centering
    \centering
    
    \includegraphics[scale=0.65]{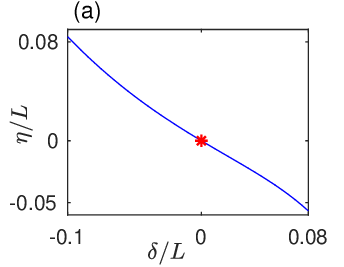} \qquad
     \includegraphics[scale=0.65]{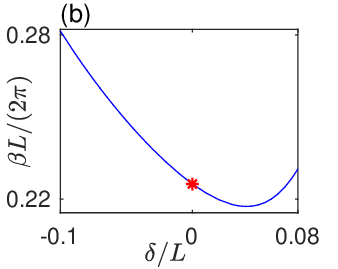} \\
     \begin{minipage}{0.5\linewidth}
      \includegraphics[scale=0.65]{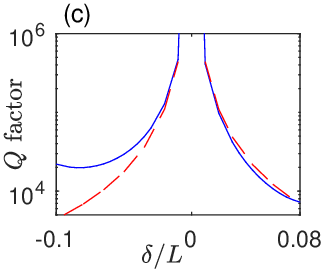}
      \includegraphics[scale=0.65]{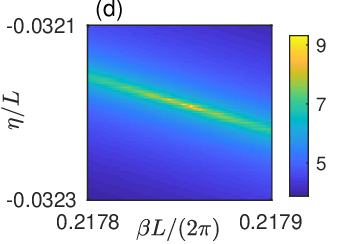}
    \end{minipage}
    \begin{minipage}{0.45\linewidth}
       \includegraphics[scale=0.6]{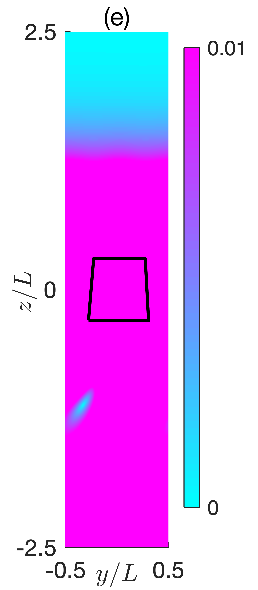}
       \end{minipage}
    \caption{Example 3: UGRs for the  perturbation given in Fig.~\ref{fig:structure}(b) near the propagating BIC. (a): $\eta$ as a function of $\delta$; (b): $\beta$ as a function of $\delta$. The red star at $\delta=0$ denotes the propagating BIC. (c): $Q$ factor of the  UGRs for various $\delta$. The dashed red line is the reference line for $Q = 43.9/ \delta^2$. (d):  $\mbox{log}(1/\tau)$ for various $\eta$ and $\beta$ with $\delta = 0.05$. (e): the wave field amplitude of the UGR at $\delta = 0.05$. The color scale is limited to $[0, 0.01]$.}
    \label{fig:example3}
\end{figure}

In the third example, we study  UGRs for the perturbation given in Fig.~\ref{fig:structure}(b) near the propagating BIC. We compute a family of UGRs for $\delta \in [-0.1, 0.08]$.  In Fig.~\ref{fig:example3}(a) and (b), the blue solid curves show  $\eta$ and $\beta$ of the UGRs as functions of $\delta$, respectively. The red star at $\delta=0$ denotes the propagating BIC given above. The $Q$ factors of the UGRs are shown in Fig.~\ref{fig:example3}(c). For $\delta = 0.05$ and $\eta = -0.0322$, we find a UGR with $\beta = 0.2179 (2\pi/L)$ and $\omega L / (2 \pi c) = 0.5785-1.93\times 10^{-5} {\sf i}$. Its wave field amplitude is shown in Fig.~\ref{fig:example3}(e). The color scale is limited to $[0, 0.01]$.  In Fig.~\ref{fig:example2}(d), we show $\mbox{log}(1/\tau)$ for resonant modes with  $\eta$ and $\delta$ near those of the UGR.

\section{Conclusions}
\label{sec:conclusions}
In summary, we developed a theory on the parametric dependence of UGRs in 2D structures with a 1D periodicity. It is shown that in the presence of a single radiation channel, a generic UGR can exist continuously with respect to a structural parameter, provided that another parameter is properly tuned. Our theory was established by constructing the UGR in the perturbed structure as a power series of the perturbation amplitude $\delta$. A key step is to show that for a specific incident wave, the diffraction solution at the complex resonant frequency of the UGR is well-defined. In our theory, the UGRs are assumed to be nondegenerate and generic, and the perturbation profile $G$ must satisfy condition (\ref{generic_cond}). Notice that a UGR is generic, if it satisfies conditions (\ref{generic1}) and (\ref{generic_cond_phi}), which are related to the {reciprocal} mode and the associated diffraction solution, respectively. Nongeneric UGRs may exhibit interesting properties that are worthy of further investigation. 

{Alternative approaches may be used to establish the parametric dependence theory. As we mentioned in Sec.~I, a UGR must satisfy two complex nonlinear equations, since the inverse scattering matrix has a zero column. 
Using the implicit function theorem, one may argue that for any sufficiently small $\delta$, these two equations can be solved yielding a real $\eta$ near 0, a real $\beta$ near $\beta_*$, and a complex $\omega$ near $\omega_*$. However,  this method could only provide qualitative information and requires the evaluation of the partial derivatives of the inverse scattering matrix.
In contrast, our method provides quantitative results with explicitly computable expansion coefficients, and reveals the precise conditions on both the UGR and the perturbation profile.}

Although we only analyzed scalar UGRs in 2D structures with 1D periodicity, our theory can be extended to vectorial UGRs satisfying the full Maxwell's equations in 3D structures with a 2D periodicity. 
{It should be mentioned that  our theory is still formal. To develop a mathematically  rigorous theory, it is necessary to establish the convergence of the power series (17)-(20).}

\section*{Acknowledgments}
The authors acknowledge support from the Natural Science Foundation of Chongqing, China (Grant No. CSTB2022NSCQ-MSX0610), the National Natural Science Foundation of China (Grant No. 12571384), and Research Grants Council of Hong Kong Special Administrative Region, China (project CityU 11305021).

\section*{Appendix}
\subsection*{Appendix A}

  We first define the reciprocal mode and the adjoint mode of a resonant mode.
  In a periodic structure with dielectric function $\epsilon$, assuming 
  the medium is homogeneous for $|z| > d$, a resonant mode $u$ with frequency $\omega$ 
  and Bloch wave number $\beta \in [-\pi/L, \pi/L]$ is a solution of the following eigenvalue problem (EVP) \cite{anne94}:
\begin{eqnarray}
\label{EVP1}
\left\{ \begin{aligned}
& \left[ \Delta + k^2 \epsilon(y,z) \right] u(y,z) = 0, & (y,z) \in \Omega_d, \\
& \mbox{Quasi-periodic in $y$ with $\beta$,} & |z| < d, \\
& \partial_z u(y, \pm d) = \pm \mathcal{T} u(y, \pm d), & |y| < \frac{L}{2},
\end{aligned}
\right.
\end{eqnarray}
where 
$\Omega_d = \{ (y,z) :  |y| < L/2,\ |z| < d \}$,
$\Delta$ is the 2D Laplacian, and  $\mathcal{T}$ is the operator 
defined by $\mathcal{T} e^{{\sf i} \beta^{(j)} y} = {\sf i} \gamma^{(j)} e^{{\sf i} \beta^{(j)} y}$ for all  integer $j$, with $\beta^{(j)}$ and $\gamma^{(j)}$ given in Eq.~(\ref{fourier}). For simplicity, we refer to $u$ as the resonant mode associated with $(\beta, \omega, \epsilon)$.

By reciprocity, if there is a resonant mode $u$ for $(\beta, \omega, \epsilon)$, then there must be another resonant mode $v$ for $(-\beta, \omega, \epsilon)$. We call $v$ the reciprocal mode of $u$. Meanwhile, corresponding to EVP~(\ref{EVP1}), there is an  adjoint eigenvalue problem
\begin{eqnarray}
\label{EVPAdjoin}
\left\{ \begin{aligned}
   & \left[  \Delta + \overline{k}^2 \overline{\epsilon}(y,z) \right] w(y,z) = 0, & (y,z) \in \Omega_d, \\
   & \mbox{ Quasi-periodic in $y$ with $\beta$,} & |z| < d, \\
   & \partial_z w (y, \pm d) = \pm \mathcal{T^{\dagger}} w(y, \pm d), & |y| < \frac{L}{2},
\end{aligned} 
\right.
\end{eqnarray}
where $\mathcal{T}^{\dagger}$ is an operator defined by $\mathcal{T}^{\dagger} e^{{\sf i} \beta^{(j)} y} = - {\sf i} \overline{\gamma}^{(j)} e^{{\sf i} \beta^{(j)} y}$ for all integer $j$, with $\overline{\gamma}^{(j)}$ being the complex conjugate of $\gamma^{(j)}$. The boundary conditions in the $z$ direction imply that adjoint modes satisfy an incoming radiation condition. One can check that the adjoint mode of $u$ for $(\beta, \omega, \epsilon)$, denoted by $u^{\dagger}$, is the complex conjugate of the reciprocal mode of $u$, i.e., $u^{\dagger} = \overline{v}$.

Suppose there exists a nondegenerate UGR $u_*$ for $(\beta_*, \omega_*, \epsilon_*)$ as defined in Sec.~\ref{sec:uni_resonance}, we now show that the diffraction problem is well posed for incident waves with $(-\beta_*, \omega_*)$ from the top. 
For the incident wave $u^{(\sf in)} = e^{-{\sf i} [\beta_* y + \gamma_* (z-d)]}$ in the region $z > d$, the diffraction problem is given by the following boundary value problem (BVP) \cite{anne94,shipman10}:
\begin{eqnarray}
\label{BVP}
\left\{ \begin{aligned}
   & \left[ \Delta  + k_*^2 \epsilon_*(y,z)\right] w(y,z) = 0, & (y,z) \in \Omega_d, \\
   & \mbox{ Quasi-periodic in $y$ with $-\beta$,} & |z| < d, \\
   & \partial_z w(y, - d) = - \mathcal{T_*} w(y, - d), & |y| < \frac{L}{2}, \\
   & \partial_z w(y,  d) =  \mathcal{T_*} w(y,  d) - 2 {\sf i} \gamma_* e^{-{\sf i} \beta_* y } , & |y| < \frac{L}{2},
\end{aligned} 
\right.
\end{eqnarray}
where $\mathcal{T}_*$ is defined analogously to $\mathcal{T}$ but with $\beta$ replaced by $-\beta_*$. Since $u_*$ is a UGR, its reciprocal mode $v_*$ is a nontrivial solution of BVP~(\ref{BVP}) without the incident wave. Furthermore, the adjoint mode of $v_*$, denoted $v_*^{\dagger}$, is $\overline{u}_*$, i.e., $v_*^{\dagger} = \overline{u}_*$. By the Fredholm alternative, BVP~(\ref{BVP}) admits a solution if and only if the incident wave is orthogonal to $v_*^{\dagger}$. More precisely, a solution exists if and only if 
$$\int_0^L \overline{v_*^{\dagger}}(y,d) e^{-{\sf i} \beta_* y } dy = \int_0^L u_*(y,d) e^{-{\sf i} \beta_* y } dy =0 .$$
This condition is automatically satisfied by the definition of the UGR in Sec.~\ref{sec:uni_resonance}.

Resonant frequencies are poles of the determinant of the scattering matrix \cite{popov86, wu22}. For a generic resonant mode, the resonant frequency is also a pole of each element of the scattering matrix. However, the existence of a diffraction solution corresponding to a UGR implies that the resonant frequency is not a pole of certain elements. Specifically, we denote the scattering matrix at $(\beta, \omega)$ by $S(\beta, \omega) = \left[ \begin{matrix} r_u &  t_d \\  t_u  & r_d  \end{matrix}\right]$, where $r_u$ and $t_u$ are the reflection and transmission coefficients for incident waves from the top,  $r_d$ and $t_d$ are those for incident waves from the bottom.  
By reciprocity \cite{wu22}, 
\begin{equation}
\label{reciprocal}
S(-\beta, \omega) = S^{\sf T}(\beta, \omega).
\end{equation}
The existence of the diffraction solution implies that
$$S(-\beta_*, \omega_*) \left[ \begin{matrix} 1  \\  0   \end{matrix}\right] = S^{\sf T}(\beta_*, \omega_*) \left[ \begin{matrix} 1  \\  0   \end{matrix}\right] = \left[ \begin{matrix} r_u(\beta_*, \omega_*) \\  t_d(\beta_*, \omega_*)   \end{matrix}\right]$$
is well defined. Thus, $(\beta_*, \omega_*)$ is not a pole of $r_u$ and $t_d$. 

If the structure lacks reflection symmetry in $y$ and $\beta_* \neq 0$, then the reciprocal mode is typically not a UGR,  and $r_d(\beta_*, \omega_*)$ and $t_u(\beta_*,  \omega_*)$ are not well defined in geneal. In that case, the diffraction problem is not well posed for incident waves with $(\beta_*, \omega_*)$ from the top. 
If the structure has reflection symmetry in $y$ or $\beta_* = 0$, then $r_d = r_u$ for all $\omega$, then $r_d(\beta_*, \omega_*)$ is well defined. Thus, the reciprocal mode is also a UGR that radiates only to the same side. In that case, the diffraction problem is also well posed for incident waves with $(\beta_*, \omega_*)$ from the top.

\subsection*{Appendix B}
The definitions of operator $\hat{\mathcal{L}}_*$  and functions $B_1$, $B_2$, $B_3$ and $C_j$ for $j \geq 1$ are
\begin{eqnarray*}
    \hat{\mathcal{L}}_* &=& \partial^2_y + \partial^2_{\hat{z}} + 2 {\sf i} \beta_* \partial_y +  k_*^2 \epsilon_* - \beta_*^2, \\
    B_1 & =& 2 \beta_* \phi_* - 2 {\sf i} \partial_y \phi_*, \\
    B_2 & = & - \epsilon_* \phi_*,  \\
    B_3 & = & - k_*^2 G \phi_*, \\
    C_j & = & V_j \phi_* + \sum_{m=1}^{j-1} \mathcal{M}_m \phi_{j-m},\\
    V_j &=& \sum_{m=1}^{j-1} \left[ \eta_m g_{j-m} G - \beta_m \beta_{j-m} \right] + g_{j-1} F, \\
    \mathcal{M}_m & = & V_m + g_m \epsilon_* + k_*^2 \eta_m G - 2 \beta_* \beta_m + 2 {\sf i} \beta_m \partial_y,
\end{eqnarray*}
where $\beta_0 = \beta_*, g_0 = k_*^2$. 

The definitions of the coefficients and the right-hand side of the linear system (\ref{system_betaj}) for $j \geq 1$ are 
\begin{eqnarray*}
    a_{1m} &=& \int_{\hat{\Omega}} \psi_*  B_m dyd{\hat{z}}, \quad  a_{2m} = \int_{\hat{\Omega}} \varphi_* B_m dyd{\hat{z}},  \quad m=1,2, 3 \\
    b_{1j} &=& \int_{\hat{\Omega}} \psi_* C_j dyd{\hat{z}}, \quad b_{2j} = \int_{\hat{\Omega}} \varphi_*  C_j dyd{\hat{z}}.
\end{eqnarray*}
From the normalization condition (\ref{normalization2}) and the orthogonal condition (\ref{orth}), we have $a_{12} = -1$ and $a_{22} = 0$, respectively. In addition, condition (\ref{real}) implies that $a_{21}$ is real.

To derive the first equation of system (\ref{system_betaj}), we multiply $\psi_*(y,\hat{z}) $ on both sides of Eq.~(\ref{eq_phij}) and integrate over domain $\hat{\Omega}_H = (-L/2, L/2) \times \Gamma_H$, where  
$$\Gamma_H = \left\{  \hat{z} \in \mathbb{C} : \ \hat{z} = \int_0^z s(\zeta) d \zeta, z < H \right\}$$ and $H > d$ is a constant. This lead to
$$
\int_{\hat{\Omega}_H} \psi_*  \hat{\mathcal{L}}_* \phi_{j} dyd{\hat{z}} = \int_{\hat{\Omega}_H} \psi_*  [\beta_j B_1 + g_j B_2 + \eta_j B_3 - C_j] dyd{\hat{z}}.
$$
The first equation of system (\ref{system_betaj}) is equivalent to $\lim\limits_{H \to +\infty} \int_{\hat{\Omega}_H} \psi_*  \hat{\mathcal{L}}_* \phi_{j} dyd{\hat{z}}  =0$. Since $\psi_*(y,\hat{z})$ satisfies $\hat{\mathcal{M}}_* \psi_*(y,\hat{z}) = 0$, where operator $\hat{\mathcal{M}}_*$ is defined as
$$ \hat{\mathcal{M}}_*  =  \partial^2_y + \partial^2_{\hat{z}} - 2 {\sf i} \beta_* \partial_y +  k_*^2 \epsilon_* - \beta_*^2, $$
we have $\int_{\hat{\Omega}_H} \phi_j(y,\hat{z}) \hat{\mathcal{M}}_* \psi_{*}(y, \hat{z}) dyd{\hat{z}} = 0$.  Using integration by parts and the periodicity in $y$, we have
\begin{eqnarray}
\label{append_1}
\nonumber    \int_{\hat{\Omega}_H} \psi_*  \hat{\mathcal{L}}_* \phi_{j} dyd{\hat{z}} &=& \int_{\hat{\Omega}_H} \psi_*  \hat{\mathcal{L}}_* \phi_{j} dyd{\hat{z}} - \int_{\hat{\Omega}_H} \phi_j  \hat{\mathcal{M}}_* \psi_{*} dyd{\hat{z}} \\ 
   & =& \int^{L/2}_{-L/2} [\psi_* \partial_z \phi_j - \phi_j \partial_z \psi_*]_{z = H} dy.
\end{eqnarray}

If the right hand-side of Eq.~(\ref{eq_phij}) decays to zero as $z \to +\infty$, then $\phi_j(y,\hat{z})$ and $\varphi_*(y,\hat{z})$ have the following asymptotic relations
\begin{equation*}
    \phi_j(y,\hat{z}) \sim d_j e^{ {\sf i} \gamma_* (z -d )}, \quad z \to +\infty,
\end{equation*}
and
\begin{equation*}
    \psi_*(y,\hat{z}) \sim b_0 e^{ {\sf i} \gamma_* (z -d )}, \quad z \to +\infty,
\end{equation*}
respectively.
Using the above two asymptotic relations in Eq.~(\ref{append_1}) and letting $H \to +\infty$, we have
$$ \int_{\hat{\Omega}} \psi_*  \hat{\mathcal{L}}_* \phi_{j} dyd{\hat{z}}  = \lim\limits_{H \to +\infty} \int_{\hat{\Omega}_H} \psi_* \hat{\mathcal{L}}_* \phi_{j} dyd{\hat{z}}  =0.  $$

The second equation of system (\ref{system_betaj}) implies that the solution of Eq.~(\ref{eq_phij}) decays to zero as $z \to +\infty$. To show this, we multiply $\varphi_*(y,\hat{z}) $ on both sides of Eq.~(\ref{eq_phij}) and integrate over domain $\hat{\Omega}_H $, and we obtain
$$
\int_{\hat{\Omega}_H} \varphi_*  \hat{\mathcal{L}}_* \phi_{j} dyd{\hat{z}} = \int_{\hat{\Omega}_H} \varphi_*  [\beta_j B_1 + g_j B_2 + \eta_j B_3 - C_j] dyd{\hat{z}}.
$$
The second equation is equivalent to $\lim\limits_{H \to +\infty} \int_{\hat{\Omega}_H} \varphi_*  \hat{\mathcal{L}}_* \phi_{j} dyd{\hat{z}}  =0$. Note that $\varphi_*(y,\hat{z})$ satisfies $\hat{\mathcal{M}}_* \varphi_*(y,\hat{z}) = 0$, following the same procedure as above, we have
\begin{eqnarray*}
\label{append_2}
 \nonumber   \int_{\hat{\Omega}_H} \varphi_*  \hat{\mathcal{L}}_* \phi_{j} dyd{\hat{z}} &=& \int_{\hat{\Omega}_H} \varphi_*  \hat{\mathcal{L}}_* \phi_{j} dyd{\hat{z}} - \int_{\hat{\Omega}_H} \phi_j  \hat{\mathcal{M}}_* \varphi_{*} dyd{\hat{z}} \\ 
   & =& \int^{L/2}_{-L/2} [\varphi_* \partial_z \phi_j - \phi_j \partial_z \varphi_*]_{z = w} dy.
\end{eqnarray*}
Using the asymptotic relations of $\varphi_*(y,\hat{z})$ (i.e. Eq.~(\ref{asymp_diff2})) and $\phi_j(y,\hat{z})$ in the above equation and letting $H \to +\infty$, we have
$$  \lim\limits_{H \to +\infty} \int_{\hat{\Omega}_H} \varphi_*  \hat{\mathcal{L}}_* \phi_{j} dyd{\hat{z}}  = 2 {\sf i} \gamma_* d_j L.  $$
Therefore, the second equation of system (\ref{system_betaj})  implies $d_j = 0$ and $\phi_j(y,\hat{z})$ decays to zero as $z \to + \infty$.


\end{document}